\documentclass[pre,showpacs]{revtex4}%
\usepackage[dvips]{graphicx}
\usepackage{bm}
\usepackage{amsmath}
\usepackage{amsfonts}
\usepackage{amssymb}%
\begin{document}
\title{Surface Roughness and Effective Stick-Slip Motion}
\author{I. V. Ponomarev}
\email{ilya@uri.edu}
\author{A. E. Meyerovich}
\email{Alexander_Meyerovich@uri.edu}
\affiliation{Department of Physics, 
University of Rhode Island, 2 Lippitt Rd.,\\
Kingston\\
RI 02881-0817 }
\date{\today }

\begin{abstract}
The effect of random surface roughness on hydrodynamics of viscous
incompressible liquid is discussed. Roughness-driven contributions to
hydrodynamic flows, energy dissipation, and friction force are calculated in a
wide range of parameters. When the hydrodynamic decay length (the viscous wave
penetration depth) is larger than the size of random surface inhomogeneities,
it is possible to replace a random rough surface by effective stick-slip
boundary conditions on a flat surface with two constants: the stick-slip
length and the renormalization of viscosity near the boundary. The stick-slip
length and the renormalization coefficient are expressed explicitly via the
correlation function of random surface inhomogeneities. The effective
stick-slip length is always negative signifying the effective slow-down of the
hydrodynamic flows by the rough surface (stick rather than slip motion). A
simple hydrodynamic model is presented as an illustration of these general
hydrodynamic results. The effective boundary parameters are analyzed
numerically for Gaussian, power-law and exponentially decaying correlators
with various indices. The maximum on the frequency dependence of the
dissipation allows one to extract the correlation radius (characteristic size)
of the surface inhomogeneities directly from, for example, experiments with
torsional quartz oscillators.

\end{abstract}
\maketitle



\section{\label{sec:sec1}Introduction}

Progress in micro- and nanotechnology requires better understanding of
boundary effects. For hydrodynamic microflows, this means better understanding
of stick-slip motion near solid walls and, in particular, information on the
dependence of the slip (or stick) length on the properties of the walls.
Despite the fact that similar issues were first raised more than a hundred
years ago \cite{max1,knu1,kun1}, the slip length remains one of the least
known transport coefficients. Traditionally, the most detailed information on
the boundary slip is available for rarefied classical
gases\cite{wel1,wil1,alb1,kog1} in applications to vacuum technology, high
altitude flights, and space research. More recently
\cite{parp1,sau1,jaf,jen1,ein1,smi1,ein2,r21}, liquid $^{3}He$ has become an
important source of information on surface slip. This is not surprising since,
in contrast to classical gases, one can easily vary the quasiparticle mean
free path in $^{3}He$ by changing temperature thus allowing experiments in a
wide range of Knudsen numbers.

The conventional theory of boundary slip assumes that the slip length is
proportional to the bulk mean free path $\mathcal{L}_{b}$, $\mathcal{L}%
_{sl}=\alpha\mathcal{L}_{b}$, and ignores small-scale surface inhomogeneities.
Obviously, this approximation is too crude. The hydrodynamic flows near the
walls strongly depend on geometry of surface inhomogeneities \cite{karn1}.
Recent analysis of slip near a model surface with periodic irregularities
demonstrated \cite{ein3} that the effective slip length $\mathcal{L}_{eff}$
contains not only the bulk component $\alpha\mathcal{L}_{b}$ but also the
contribution from the averaged surface curvature $R$, $\mathcal{L}_{eff}%
^{-1}=\alpha^{-1}\mathcal{L}_{b}^{-1}-R^{-1} $. An application of the
corresponding boundary condition to several types of curved walls \cite{liu1}
resulted in an interesting expression for an effective slip length which
could, under certain circumstances, be equivalent to large-scale surface
roughness. However, the results \cite{ein3,liu1} were obtained for few special
types of regular surface inhomogeneities only. In the case of micro- and
nanoscale defects, it is more realistically to suggest that surfaces have
\textit{random} corrugation. What is more, in some cases, especially in the
hydrodynamic limit $\mathcal{L}_{b}\rightarrow0$, it is not clear how to use
the effective boundary parameters of Refs. \cite{ein3,liu1}.

Below we derive an effective stick-slip boundary condition which would
reproduce hydrodynamic flows with $\mathcal{L}_{b}=0$ near rough walls with
small-scale \emph{random }inhomogeneities. Since the hydrodynamic calculations
near inhomogeneous walls are extremely complicated \cite{karn1}, it is highly
desirable to map this problem onto the system with simple flat surface
geometry with some effective boundary condition. This boundary condition
should contain information about geometrical and statistical properties of the
real corrugated surface and ensure a proper behavior of hydrodynamic
variables. The derivation of this simple boundary condition is the main goal
of the paper. Below we show that this boundary condition contains two
effective parameters: the effective stick-slip length and renormalized
viscosity. We also demonstrate that the results for attenuation in torsional
oscillator experiments can provide valuable information about the statistical
type of surface inhomogeneities and give the values of the main geometrical
parameters of surface roughness.

In the next Section, we present the main hydrodynamic equations and find the
stream function in systems with random rough walls (details of the derivation
are given in Appendix A). Comparison of the hydrodynamic results with those
for the stick-slip motion allow us to get the expression for the effective
stick-slip parameters in Section III. For clarification of the physical
meaning of the parameters in the somewhat unexpected effective boundary
condition, we present a simple hydrodynamic model for a boundary layer in
Appendix B. Section IV contains analytical and numerical results for surfaces
with various statistical types of inhomogeneities. Summary and conclusions are
presented in Section V.

\section{Hydrodynamic flows along rough walls}

To determine an effective slip/stick length, one has to solve an appropriate
hydrodynamic problem with a boundary condition on a random rough wall and to
compare the results with those for a similar problem with a slip boundary
condition on a smooth wall. Several ''typical''\ hydrodynamic problems
\cite{LL1} have been generalized recently in order to cover boundaries with
slight roughness \cite{urb1,urb2,urb3,chu1,chu1b}. For our purposes, the most
appropriate problem is the problem of hydrodynamic flows excited by tangential
oscillations of a rough wall. The advantages are the convenience of the
experimental setup with a standard transverse oscillator, a choice of several
observables such as hydrodynamic velocity and two components of the shear
impedance, and the presence of an extra variable - frequency $\omega$ - that
allows one to vary the ratio of the hydrodynamic decay length to the size of
wall inhomogeneities. Since this problem has already been studied in Ref.
\cite{urb1}, though by a different method, we will only briefly outline our
hydrodynamic formalism in Appendix A and present some additional results.

\begin{figure}[ptb]
\includegraphics{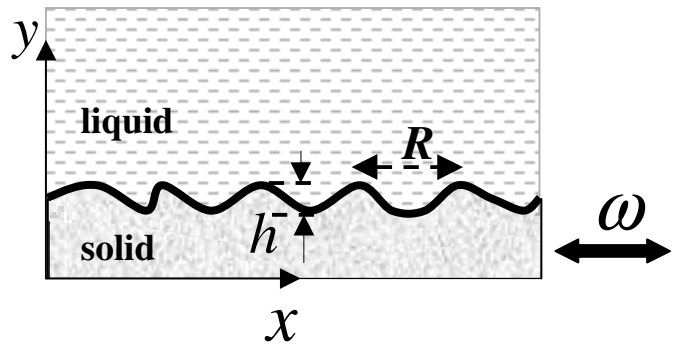}\caption{General geometry of the model.}%
\label{fig1}%
\end{figure}
We consider semi-infinite viscous fluid restricted by a rough solid wall. For
simplicity, roughness is assumed one-dimensional with profile described by a
random function $Y=\Xi\left(  X\right)  $ with the zero mean value,
$\left\langle \Xi\left(  X\right)  \right\rangle =0.$ The wall is homogeneous
in $Z$-direction (see Fig.~\ref{fig1}). This inhomogeneous surface is
characterized by two length parameters: the average amplitude $h$ and
correlation radius (size) $R$ of surface inhomogeneities. We consider the case
of slight roughness,
\begin{equation}
\epsilon=h/R\ll1.\label{ee13}%
\end{equation}
In other situations, any general description of hydrodynamic flows near rough
walls is virtually impossible.

The wall oscillates in $X$-direction with the velocity%

\begin{equation}
\mathbf{U}\left(  t\right)  =\mathbf{e}_{x}u_{0}\cos\left(  \omega t\right)
\label{eq1}%
\end{equation}
The hydrodynamic boundary condition is the condition of zero velocity
$\mathbf{V}$\ on the wall in the reference frame in which the wall is at
rest:
\begin{equation}
\mathbf{V}\left(  X-\int\mathbf{U}\left(  t\right)  dt,Y=\Xi\left(
X-\int\mathbf{U}\left(  t\right)  dt\right)  \right)  =\mathbf{0},\label{b1}%
\end{equation}

Two important hydrodynamic length scales are the decay length (or the viscous
wave penetration depth), $\delta$, and the amplitude of the boundary
oscillations, $a,$
\begin{equation}
\delta=\sqrt{2\nu/\omega},\;a=u_{0}/\omega,\label{ee12}%
\end{equation}
where $\nu=\eta/\rho$ is the kinematic viscosity.

It is convenient to choose $h,R$ and the amplitude of the wall velocity
$u_{0}$ as the scaling parameters and introduce dimensionless variables as%

\begin{equation}
\mathbf{v}=\mathbf{V}/u_{0},\;x=X/R,\;y=Y/R,\;\xi(x)=\Xi(X/R)/h.\label{e10}%
\end{equation}
When the fluctuations of $\xi(x)$ are statistically independent and the higher
momenta can be expressed through the second one, the random surface roughness
is actually described not by the unknown random function $\xi(x)$ with the
zero average, but by the correlation function $\zeta\left(  x\right)  $:%

\begin{align}
\zeta\left(  x\right)   &  \equiv\left\langle \xi\left(  x_{1}\right)
\xi\left(  x_{1}+x\right)  \right\rangle =\frac{1}{A}\int_{-\infty}^{\infty
}\xi\left(  x_{1}\right)  \xi\left(  x_{1}+x\right)  dx_{1},\label{eq2}\\
\left\langle \xi\left(  k_{x}\right)  \xi\left(  k_{x}^{\prime}\right)
\right\rangle  &  =2\pi\delta\left(  k_{x}+k_{x}^{\prime}\right)  \zeta\left(
k_{x}\right)  ,\nonumber
\end{align}
where $A$ is dimensionless flat surface area of the wall. Experimentally the
correlation functions $\zeta\left(  x\right)  $ [or its Fourier image, also
called the power spectrum, $\zeta\left(  k_{x}\right)  $] can exhibit
different types of long-range behavior and assume various forms \cite{Ogil}.
Particular examples of the surface correlators are analyzed in Section IV.
Note, that in our dimensionless notations $\left(  \ref{e10}\right)  $ the
correlation radius of the surface inhomogeneities is equal to $1$.

The liquid is considered incompressible, $\operatorname{div}\mathbf{v=}0$. In
variables $\left(  \ref{e10}\right)  $, the dimensionless Navier-Stokes
equation can be written as
\begin{equation}
\frac{1}{\omega_{0}}\frac{\partial\operatorname*{rot}\mathbf{v}}{\partial
t}-\mathbf{\nabla}^{2}\operatorname{rot}\mathbf{v}=\Re\left[  \left(
\operatorname{rot}\mathbf{v\nabla}\right)  \mathbf{v-}\left(  \mathbf{v\nabla
}\right)  \operatorname{rot}\mathbf{v}\right] \label{e11}%
\end{equation}
where the characteristic frequency $\omega_{0}$ and the Reynolds number $\Re
$\textit{\ }are
\begin{equation}
\omega_{0}=\frac{\nu}{R^{2}},\text{\ }\Re=\frac{u_{0}R}{\nu}\equiv\frac{a}%
{R}\frac{\omega}{\omega_{0}}\label{e12}%
\end{equation}
(the inverse frequency parameter $\omega_{0}^{-1}$ is often called the
diffusion time of vorticity) \ Since the first term in Eq. $\left(
\ref{e11}\right)  $ has an order of $\left(  \omega/\omega_{0}\right)
\operatorname{rot}\mathbf{v,}$ the hydrodynamic flows are characterized by the
dimensionless parameter
\begin{equation}
\Lambda=\sqrt{\omega/\omega_{0}}=\sqrt{2}R/\delta\label{eq12}%
\end{equation}
which describes the ratio of the size of inhomogeneities $R$ to the
hydrodynamic decay length $\delta$. Two dimensionless parameters, $\epsilon$
and $\Lambda$, are the main parameters of the problem.

Below we consider the linearized Navier-Stokes equation without the nonlinear
term in \textit{r.h.s.} of Eq. $\left(  \ref{e11}\right)  $. For small
frequencies, $\omega/\omega_{0}\ll1,$ this linearization is justified for
small Reynolds numbers $\Re\ll1$. In the opposite limit of high frequencies,
$\omega/\omega_{0}\gg1,$ this requires smallness of the amplitude of
oscillations $a$ in comparison with the tangential size of surface
inhomogeneities $R$ at arbitrary Reynolds numbers $\Re$, $a/R\ll1$ \cite{LL1}.
The linearized Eq. $\left(  \ref{e11}\right)  $\ for $\operatorname{rot}%
\mathbf{v}$\ can be, as usual, rewritten as the fourth order differential
equation for the scalar stream function $\psi\left(  x,y\right)  $,
\begin{equation}
v_{x}=\frac{\partial\psi}{\partial y},\;v_{y}=-\frac{\partial\psi}{\partial
x}.\label{b2}%
\end{equation}
In our problem, all hydrodynamic variables contain harmonic time dependence.
After the transformation to the coordinate frame oscillating with the wall,
the hydrodynamic equations and boundary conditions for the stream function
acquire the form
\begin{align}
-i\Lambda^{2}\nabla^{2}\psi-\nabla^{4}\psi &  =0,\label{eq7}\\
\frac{\partial\psi\left(  x,\epsilon\xi\left(  x\right)  \right)  }{\partial
y}  &  =1,\ \frac{\partial\psi\left(  x,\epsilon\xi\left(  x\right)  \right)
}{\partial x}=0,\label{eq8}\\
\psi(x,\infty)  &  =const.\label{eq9}%
\end{align}

The solution of the linearized Navier-Stokes equation $\left(  \ref{eq7}%
\right)  -\left(  \ref{eq9}\right)  $ is quite difficult because the boundary
condition $\left(  \ref{eq8}\right)  $\ involves the rough wall with random
inhomogeneities. Using a coordinate transformation $y\rightarrow y-\xi\left(
x\right)  ,$ we can reduce the Navier-Stokes equation to an equivalent
equation with the boundary condition on the perfect flat wall. However, this
new equation, as a result of the transformation-driven change in derivatives,
acquires several additional terms $\widehat{V}\psi$ that involve the
combinations of derivatives of $\psi$\ and the random function $\xi\left(
x\right)  $. To deal with these terms, we find the explicit form of the
Green's function with the proper boundary condition. Then the problem reduces
to a rather transparent integral equation
\begin{equation}
\psi\left(  k_{x},y\right)  =\psi_{inh}\left(  k_{x},y\right)  +\int
\limits_{0}^{\infty}dy^{\prime}\,G(k_{x}\mathbf{,}y,y^{\prime})\int
\limits_{-\infty}^{\infty}\frac{dk_{x}^{\prime}}{2\pi}\widehat{V}\left(
k_{x}-k_{x}^{\prime},y^{\prime}\right)  \psi\left(  k_{x}^{\prime},y^{\prime
}\right)  .\label{ineq1}%
\end{equation}
This procedure and the explicit expressions for the unperturbed inhomogeneous
solution $\psi_{inh}\left(  k_{x},y\right)  ,$ the perturbation\textbf{\ }%
$\widehat{V}$, and the Green's function are given in Appendix A. In some
sense,\ we shifted the difficulty from the boundary condition to the bulk
equations with random sources of the special form. Note, that Eq.
(\ref{ineq1}) is still exact and, in principle, could be solved without the
perturbation theory. The explicit form of Green's function is such that one
can extract the main part of the solution in the closed form. Another possible
approach to Eq. (\ref{ineq1}) is to apply the Wiener-Hermite functional
expansion \cite{whe,eft1}.

Here we solve Eq. (\ref{ineq1}) by iterations as an expansion in the small
parameter$:$ $\psi=\psi_{0}+\epsilon\psi_{1}+\epsilon^{2}\psi_{2}+\ldots$ The
first three terms for the stream function have the following forms
\begin{align}
\psi_{0}\left(  k_{x},y\right)   &  =\frac{2\pi}{i\lambda}\delta\left(
k_{x}\right)  \exp\left(  i\lambda y\right)  ,\label{psi0}\\
\psi_{1}\left(  k_{x,}y\right)   &  =\xi\left(  k_{x}\right)  \left[
e^{i\lambda y}+\frac{i\lambda}{s_{2}-s_{1}}\left(  e^{s_{1}y}-e^{s_{2}%
y}\right)  \right]  ,\label{psi1}\\
\left\langle \psi_{2}\left(  k_{x},y\right)  \right\rangle  &  =\delta\left(
k_{x}\right)  \int\limits_{-\infty}^{\infty}dk_{x}^{\prime}\,\zeta
(k_{x}^{\prime})\left[  \frac{s_{1}+s_{2}}{i\lambda}\left(  i\lambda
e^{i\lambda y}+s_{1}e^{s_{1}y}-s_{2}e^{s_{2}y}\right)  \right]  ,\label{psi2}%
\end{align}
where we exclude uninteresting constant terms and
\begin{align}
s_{1}  &  =-\left|  k_{x}\right|  ,\ s_{2}=\sqrt{k_{x}^{2}-i\Lambda^{2}}%
\equiv-\alpha+i\beta,\nonumber\\
\alpha,\beta &  =\frac{1}{\sqrt{2}}\sqrt{\sqrt{k_{x}^{4}+\Lambda^{4}}\pm
k_{x}^{2}}\geq0,\nonumber\\
\lambda &  =e^{i\pi/4}\Lambda.\label{s12}%
\end{align}
Since for further calculations we need only the expression for $\psi_{2}$
which is averaged over the random surface inhomogeneities, Eq. (\ref{psi2})
gives only the compact expression for $\left\langle \psi_{2}\left(
k_{x},y\right)  \right\rangle $.

These expressions for the stream function provide the roughness-driven
corrections for the velocity and rate of energy dissipation (see Appendix A):%

\begin{align}
\left\langle v_{x}\right\rangle  &  =\operatorname{Re}\left\{  e^{i\left(
\lambda y-\omega t\right)  }\left[  1+i\lambda\epsilon^{2}\ell_{1}\right]
\right\}  ,\ \left\langle v_{y}\right\rangle =0,\label{eq18}\\
\ell_{1}  &  =\int\limits_{0}^{\infty}\frac{dk_{x}}{\pi}\zeta\left(
k_{x}\right)  \left\{  s_{1}+s_{2}-i\lambda/2\right\} \label{eq18b}\\
Q  &  =-\frac{\eta u_{0}^{2}}{2R}\frac{\Lambda}{\sqrt{2}}\left[
1+\epsilon^{2}\Lambda^{2}\ell_{2}\right]  ,\label{eq18c}\\
\ell_{2}  &  =\int\limits_{0}^{\infty}\frac{dt}{\pi}\zeta\left(
t\Lambda\right)  \phi\left(  t\right)  ,\ \phi\left(  t\right)  =1-\sqrt
{\sqrt{1+t^{4}}-t^{2}}.\label{eq18d}%
\end{align}
The equation for the energy dissipation is averaged over both the surface
roughness and the period of oscillations. This expression is similar to the
result of Ref. \cite{urb1} obtained with the help of the Rayleigh perturbation method.

Stream function also allows one to find corrections to the roughness-driven
friction force. These calculations should be done more carefully than for
standard flat geometry: the friction force is parallel to the actual surface
and, in the case of the oscillating \emph{rough} wall, has both components
$F_{x}$ and $F_{y}$. One should also take into account the $y$-component of
velocity, which is absent in the case of flat geometry. Straightforward
calculation for the averaged square of absolute value of dimensionless
friction force give
\begin{equation}
\mathbf{F=}\frac{\eta u_{0}}{R}\mathbf{f},\ \left\langle \overline{f^{2}%
}\right\rangle =\frac{\Lambda^{2}}{2}\left[  1+\frac{\epsilon^{2}\Lambda^{2}%
}{\pi}\int\limits_{0}^{\infty}dt\zeta(\Lambda t)\phi^{2}\left(  t\right)
\right]  .\label{eq23}%
\end{equation}
This expression is different from a simple experimental definition of the
effective friction force $F_{eff}=-Q/u_{0}$

At low frequencies (large decay lengths, $\Lambda\ll1$), Eqs. $\left(
\ref{eq18}\right)  ,\left(  \ref{eq18c}\right)  $ expressions for parameters
$\ell_{1,2}$ reduce to%
\begin{align}
\ell_{1}  &  =-2\int\limits_{0}^{\infty}\frac{dk_{x}}{\pi}\zeta\left(
k_{x}\right)  k_{x}+O\left(  \Lambda\right)  ,\label{l1}\\
\ell_{2}  &  =1+O\left(  \Lambda\ln\Lambda\right)  ,\label{l22}%
\end{align}
and the equations for the velocity and attenuation acquire the following
form:
\begin{align}
\left\langle v_{x}\right\rangle  &  =\operatorname{Re}\left\{  e^{i\left(
\lambda y-\omega t\right)  }\left[  1-\Lambda e^{i3\pi/4}\epsilon^{2}\left(
2\int\limits_{0}^{\infty}\frac{dk_{x}}{\pi}\zeta\left(  k_{x}\right)
k_{x}+O\left(  \Lambda\right)  \right)  \right]  \right\}  ,\label{fr1}\\
Q  &  =-\frac{\eta u_{0}^{2}}{2R}\frac{\Lambda}{\sqrt{2}}\left[  1+\Lambda
^{2}\epsilon^{2}\left(  1+O\left(  \Lambda\ln\Lambda\right)  \right)  \right]
.\label{fr22}%
\end{align}
The fact that the main term in $\ell_{2}$\ is equal to 1 is due to our choice
of the normalization of the correlation function in Eq. $\left(
\ref{e10}\right)  $ as $\zeta\left(  x=0\right)  =1$ (see also Section IV).

In the opposite limit of high frequencies $\Lambda\gg1$,
\begin{equation}
Q\approx-\frac{\eta u_{0}^{2}}{2R}\frac{\Lambda}{\sqrt{2}}\left[
1+\frac{\epsilon^{2}}{2}\int\limits_{0}^{\infty}\frac{dk_{x}}{\pi}\zeta\left(
k_{x}\right)  k_{x}^{2}\right]  \equiv Q_{0}\left[  1+\frac{\epsilon^{2}}%
{2}\left\langle \xi^{\prime2}\right\rangle \right]  .\label{lf1}%
\end{equation}
This result has a simple physical explanation. In this limit, the decay length
is much smaller than correlation radius (size) of the wall inhomogeneities
$R$. As a result, the dissipation occurs in a very narrow layer near the wall
within which the wall can be considered as almost flat. Then the correction to
dissipation stems simply from the increase in the surface area relatively to
the flat boundary
\begin{equation}
Q\approx-\frac{\eta u_{0}^{2}}{2R}\frac{\Lambda}{\sqrt{2}}\frac{1}{L^{2}}\oint
dA=-\frac{\eta u_{0}^{2}}{2R}\frac{\Lambda}{\sqrt{2}}\frac{1}{L}\int
\sqrt{1+\epsilon^{2}\xi^{\prime2}}dx\label{lf2}%
\end{equation}
Equation $\left(  \ref{lf1}\right)  $ is simply the combination of the first
two terms in the Taylor expansion of Eq. $\left(  \ref{lf2}\right)  $ in small
$\epsilon.$

In principle, it is possible to slightly modify our problem by considering a
torsional quartz crystal oscillator with density $\rho_{s}$,\ thickness $d$.
If such a resonator has a rough solid-fluid interface, the frequency shift
$\delta\omega$ of the resonance frequency $\Omega_{0}$\ acquires an additional
roughness-driven component which can be described within the above formalism
and should be given by the similar equations. Such a frequency shift for a
transverse oscillator is \cite{urb1}
\[
\delta\omega=-\frac{\eta}{\sqrt{2}}\frac{\Lambda}{R}\frac{1}{\rho_{s}%
d}\left\{  1+\epsilon^{2}\Lambda\int_{0}^{\infty}\frac{dk_{x}}{\pi}%
\zeta\left(  k_{x}\right)  \left[  \sqrt{\sqrt{k_{x}^{4}+\Lambda^{4}}%
+k_{x}^{2}}-\Lambda+\sqrt{2}k_{x}\right]  \right\}  .
\]
We do not want to dwell on this issue; our interest in focused mainly on the
roughness-driven corrections to the hydrodynamic flows and dissipation.

\section{Effective stick-slip boundary conditions}

The main aim of the paper is to find when and to what extent flows near random
rough surface are equivalent to stick-slip motion with some effective
stick-slip boundary conditions near flat surfaces,
\begin{equation}
\operatorname{Re}\left\{  v_{x}\left(  x,0,t\right)  -\frac{\mathcal{L}_{eff}%
}{R}\frac{\partial v_{x}\left(  x,0,t\right)  }{\partial y}\right\}
=\operatorname{Re}\left(  e^{-i\omega t}\right) \label{sl1}%
\end{equation}
where the effective stick-slip length $\mathcal{L}_{eff}$, in order to
simplify the applications of the results, is introduced with the proper
dimensionality of length while all other variables are still dimensionless,
Eq. $\left(  \ref{e10}\right)  $. With this boundary condition on a flat wall,
the velocity field is
\begin{equation}
v_{x}\left(  y,t\right)  =\operatorname{Re}\left[  \frac{e^{i\left(  \lambda
y-\omega t\right)  }}{1-e^{i3\pi/4}\Lambda\mathcal{L}_{eff}/R}\right]
\label{sl2}%
\end{equation}
Since the roughness-generated corrections for velocity are small, the
comparison between Eq.$\left(  \ref{sl2}\right)  $ and Eqs.$\left(
\ref{eq18}\right)  ,\left(  \ref{fr1}\right)  $ is possible only when
$\Lambda\mathcal{L}_{eff}/R\ll1$, \textit{i.e.}, only for relatively large
decay lengths (low frequencies),
\begin{equation}
v_{x}\left(  y,t\right)  \approx\operatorname{Re}\left[  e^{i\left(  \lambda
y-\omega t\right)  }\left(  1+e^{i3\pi/4}\Lambda\mathcal{L}_{eff}/R\right)
\right] \label{sl3}%
\end{equation}
In this case, the comparison with the roughness-driven correction for the
velocity at low frequencies, Eq. $\left(  \ref{fr1}\right)  $, yields the
following simple expression for the effective stick-slip length $\mathcal{L}%
_{eff}=R\epsilon^{2}\ell_{1}$:
\begin{equation}
\mathcal{L}_{eff}=-2\frac{h^{2}}{R}\int\limits_{0}^{\infty}\frac{dk_{x}}{\pi
}\zeta\left(  k_{x}\right)  k_{x}\label{sl4}%
\end{equation}
The negative sign in Eq. $\left(  \ref{sl4}\right)  $\ means that the rough
boundary roughness causes effective slow-down of the liquid, \textit{i.e.,}
that the coefficient $\mathcal{L}_{eff}$ $\left(  \ref{sl4}\right)  $ is the
stick length rather than the slip length. In other words, there is an
additional roughness-induced friction.

The condition $\Lambda\mathcal{L}_{eff}/R\ll1$, when one can replace the rough
wall by a stick-slip boundary condition is equivalent to\
\begin{equation}
\frac{\Lambda\mathcal{L}_{eff}}{R}\sim\Lambda\epsilon^{2}\sim\frac{h^{2}%
}{R\delta}\ll1\label{cond}%
\end{equation}

Surprisingly, the effective boundary condition $\left(  \ref{sl1}\right)
$,$\left(  \ref{sl4}\right)  ,$ taken by itself, cannot emulate the
roughness-driven attenuation $\left(  \ref{eq18c}\right)  $. The reason is the
presence normal flows near the boundary, $v_{y}\left(  x\right)  $, which are
completely absent within the effective stick-slip description $\left(
\ref{sl1}\right)  ,\left(  \ref{sl2}\right)  $ in which $v_{y}=0.$ The
attempts to modify the boundary condition $\left(  \ref{sl1}\right)  $ so that
to reproduce both the velocity and attenuation correctly by, for example,
introducing a two-component or complex stick-slip length, fail. In order to
emulate the correct behavior of liquid near a rough wall, one has not only to
introduce the stick-slip length $\left(  \ref{sl1}\right)  $,$\left(
\ref{sl4}\right)  ,$ but also to renormalize the viscosity near the wall as
\begin{equation}
\eta_{eff}\left(  y\right)  =\eta\left[  1+\beta\delta\left(  y\right)
\right]  ,\label{sl5}%
\end{equation}
where renormalization parameter $\beta$ is equal at small $\Lambda$ to%
\begin{equation}
\beta_{\,}\approx2\left[  \frac{\mathcal{L}_{eff}}{R}+\frac{\Lambda}{\sqrt{2}%
}\frac{h^{2}}{R^{2}}\right]  .\label{sl6}%
\end{equation}

The effective boundary conditions $\left(  \ref{sl1}\right)  $,$\left(
\ref{sl4}\right)  ,\left(  \ref{sl5}\right)  $,$\left(  \ref{sl6}\right)  $
are the main result of this paper. These conditions allow one to replace the
random rough boundary by an equivalent problem with the flat boundary and the
effective stick-slip length and renormalized viscosity. The necessity of the
renormalization of the viscosity means that the rough surface slows the flow
down and changes the attenuation. Usually, the slip boundary condition is
understood in terms of the existence of a peculiar thin slip boundary layer
with the thickness of the order of the mean free path and with \ the
properties that are somewhat different from the rest of the liquid. In the
case of the rough walls, one should not only introduce the effective
stick-slip layer with the thickness that is determined by $\delta$ and $R$,
but also to renormalize the viscosity in this layer explicitly. A simple
physical model that clarifies the meaning of the effective parameters is given
in Appendix B.

\section{Comparison for different types of random inhomogeneities}

In this Section we address the question whether it is possible to extract
information on the properties of the rough surface from the frequency
dependence of attenuation of transverse oscillations. Statistical properties
of the random surface are described by the correlation function of surface
inhomogeneities, $\Xi\left(  X\right)  =h^{2}\zeta(x),\ x=X/R$, Eq. $\left(
\ref{eq2}\right)  .$ Experimentally, the correlation function can exhibit
different types of long-range behavior and can assume various forms
\cite{Ogil}.

Three broad classes of the correlation functions $\zeta\left(  x\right)  $ and
their Fourier images $\zeta\left(  k\right)  $ (the so-called power density
spectral function, or power spectra) are summarized in Table~\ref{tab:table1}.
For better comparison, all the correlators are normalized in the same way,
$\zeta\left(  x=0\right)  =1$. Note, that this normalization differs from the
one used in Ref. \cite{pon1} for conductivity of ultrathin films: the natural
reference point for the conductivity was its value in the limit $kR\rightarrow
0$ and all the correlation functions in Ref. \cite{pon1} have been normalized
using $\zeta\left(  k=0\right)  =1$. For the hydrodynamic problem in this
paper, the normalization $\zeta\left(  x=0\right)  =1$ provides a better reference.

\begin{table}[ptb]
\caption{The position of the maximum of the funcion (\ref{loss1}) and the
value of stick-slip length Eq. (\ref{z3}) for different types of the surface
correlation function.}%
\label{tab:table1}%
\begin{ruledtabular}
\begin{tabular}{cccccc}
$\sharp$ & Correlator type & Form, $\zeta\left( x\right) $ & Fourier
image, $\zeta\left( k\right) $ & $\Lambda_{\max}$ & $-\ell_{1}\left(
\Lambda\ll1\right) $ \\
\hline
1. & Gaussian & $\exp\left( -x^{2}\right) $ & $\sqrt{\pi}\exp\left(
-k^{2}/4\right) $ & 1.293 & $4/\sqrt{\pi}$ \\
2 & Power-law & $\left( 1+x^{2}\right) ^{-\left( \mu+1/2\right) }$ & $\frac{%
\sqrt{\pi}}{2^{\mu-1}\Gamma\left( \mu+1/2\right) }\left\vert
k\right\vert^{\mu}K_{\mu}\left( \left\vert k\right\vert\right) $ &  & $%
\frac{4}{\sqrt{\pi}}\frac{\Gamma\left( \mu+1\right) }{\Gamma\left( \mu
+1/2\right) }$ \\
2a & $\mu=1/2$: Lorentzian & $\left( 1+x^{2}\right) ^{-1}$ & $\pi\exp
\left( -\left\vert k\right\vert\right) $ & 1.320 & $2$ \\
2b & $\mu=3/2$: Staras & $\left( 1+x^{2}\right) ^{-2}$ & $\frac{\pi}{2}%
\left( 1+\left\vert k\right\vert\right) \exp\left( -\left\vert
k\right\vert\right) $ & 1.825 & $3$ \\
3 & Power-law Fourier image & $\frac{1}{2^{\nu-1}\Gamma\left( \nu\right) }%
\left\vert x\right\vert^{\nu}K_{\nu}\left( \left\vert x\right\vert
\right) $ & $2\sqrt{\pi}\frac{\Gamma\left( \nu+1/2\right) }{\Gamma\left(
\nu\right) }\left( 1+k^{2}\right) ^{-\left\{ \nu+1/2\right\} }$ &  & $%
\frac{2}{\sqrt{\pi}}\frac{\Gamma\left( \nu+1/2\right) }{\Gamma\left( \nu
\right) }\frac{1}{\nu-1/2}$ \\
3a & $\nu=1/2$: exponential & $\exp\left( -\left\vert x\right\vert\right)
$ & $2\left( 1+k^{2}\right) ^{-1}$ & no $\max$ & $\infty$ \\
3b & $\nu=3/2$ & $\left( 1+\left\vert x\right\vert\right) \exp\left(
-\left\vert x\right\vert\right) $ & $4\left( 1+k^{2}\right) ^{-2}$ & 1.238
& $4/\pi$ \\
\end{tabular}
\end{ruledtabular}
\end{table}

The most commonly used correlation function, namely, the Gaussian\ correlator,
is listed first. The next class of the correlation functions covers power-law
correlators with the exponentially decaying Fourier images (power spectra),
$\left|  k\right|  ^{\mu}K_{\mu}\left(  \left|  k\right|  \right)  $. Here,
the most widely used are the Lorentzian correlator (index $\mu=1/2$) and the
Staras correlator ($\mu=3/2$). The third class of the correlation functions
includes the conjugate correlators, namely, the exponentially decaying
correlators with the power-law spectral function $\zeta\left(  k\right)  $. In
our dimensionless notations, Eq. $\left(  \ref{e10}\right)  $, all the
correlators have the correlation radius equal to one.

The most convenient observable is the frequency dependence of the relative
attenuation, Eq.$\left(  \ref{eq18c}\right)  $:
\begin{align}
\Delta\Gamma\left(  \Lambda\right)   &  =\frac{\Delta Q}{\Lambda\epsilon
^{2}Q_{0}}\equiv\Lambda\int\limits_{0}^{\infty}\frac{dt}{\pi}\zeta\left(
t\Lambda\right)  \phi\left(  t\right)  ,\label{loss1}\\
\phi\left(  t\right)   &  =1-\sqrt{\sqrt{1+t^{4}}-t^{2}}.\label{l2}%
\end{align}
In the limits $t\ll1$ and $t\gg1$, the function $\phi\left(  t\right)  $ has
the following asymptotic expansions:
\begin{equation}
\phi\left(  t\right)  \approx\left\{
\begin{array}
[c]{cc}%
t^{2}/2-t^{4}/8, & t\ll1\\
1-1/\sqrt{2}t, & t\gg1.
\end{array}
\right. \label{phiapr}%
\end{equation}
Note, that the piecewise continues function, defined by the expressions in Eq.
$\left(  \ref{phiapr}\right)  $ connected at the point $t=\sqrt{2},$ gives a
good approximation for $\phi\left(  t\right)  $ in the whole range of $t$.
This can be useful in simple approximations of the integral $\left(
\ref{loss1}\right)  $. The integral $\left(  \ref{loss1}\right)  $ can be
conveniently split into two parts, $\Delta\Gamma_{1}$\ and $\Delta\Gamma_{2}$,
which correspond to the contributions from small and large $t$.

In the laminar limit, $\Lambda\ll1,$ the main contribution comes from large
$t:$
\begin{equation}
\Delta\Gamma\sim\Delta\Gamma_{2}=\Lambda-\Lambda^{2}\ln\left(  \Lambda\right)
/\sqrt{2}+O\left(  \Lambda^{2}\right)  .\label{smlam}%
\end{equation}
The first two terms in this expression are the same for the correlators of all
types. Therefore, in the low-frequency limit with large decay length it is
impossible to distinguish statistical properties of different surfaces. The
physical reason is obvious: large-scale attenuation processes on the scale of
decay length $\delta$ are not very sensitive to the details of surface
inhomogeneities with the size $R\ll\delta$.

The situation is different in the opposite case of large $\Lambda.$ In this
limit for Gaussian and power-law correlators with the exponential power
spectra (types 1 and 2 in the Table~\ref{tab:table1}), the contribution from
large $t$ to the integral $\left(  \ref{loss1}\right)  $ is exponentially
small. An estimate of the contribution from small $k$ yields
\begin{equation}
\Delta\Gamma\sim\Delta\Gamma_{1}\approx\int_{0}^{\infty}\frac{dk}{\pi}%
\frac{k^{2}\zeta\left(  k\right)  }{2\Lambda}=-\frac{1}{2\Lambda}\left.
\frac{d^{2}\zeta\left(  x\right)  }{dx^{2}}\right\vert _{x=0}=\frac{a}%
{\Lambda},\label{llexp1}%
\end{equation}
where $a=1$ for the Gaussian correlator and $a=\mu+1/2$ for the power-law correlators.

For the correlators with the power-law power spectrum (correlators of the type
3 in the Table~\ref{tab:table1}), the contribution from large $t$,
$\Delta\Gamma_{2}$, is
\[
\Delta\Gamma_{2}\sim\Lambda\int_{0}^{1/\Lambda}\frac{t^{2\nu-1}dt}{\left(
1+t^{2}\right)  ^{\nu+1/2}}\sim\Lambda^{1-2\nu}.
\]
The contribution from small $t$, $\Delta\Gamma_{1}$, strongly depends on the
value of the exponent $\nu.$ If $0<\nu<1,$ then the value of $\Delta\Gamma
_{1}$ is determined by the upper limit of the corresponding part of the
integral and it is also proportional to $\Lambda^{1-2\nu}.$ If $\nu>1, $ then
the first terms in the Taylor expansion for $\phi\left(  t\right)  $ yields a
convergent integral proportional to $\Lambda^{-1}$, while the rest gives the
terms with the smaller exponent $\Lambda^{1-2\nu}:$
\[
\Delta\Gamma_{1}\sim\int_{0}^{\Lambda}\frac{dk}{\left(  1+k^{2}\right)
^{\left\{  \nu+1/2\right\}  }}\left[  \frac{k^{2}}{2\Lambda}-\frac{k^{4}%
}{8\Lambda^{3}}+...\right]  \sim\frac{1}{\Lambda}+O\left(  \frac{1}%
{\Lambda^{2\nu-1}}\right)  .
\]
Thus, the energy dissipation rate for the correlators with the power-law power
spectrum is determined by the value of the index $\nu:$%
\begin{equation}
\Delta\Gamma\left(  \Lambda,\nu\right)  \sim\left\{
\begin{array}
[c]{cc}%
\Lambda^{1-2\nu}, & 0<\nu<1\text{,}\\
\Lambda^{-1}, & \nu>1\text{.}%
\end{array}
\right. \label{llpower}%
\end{equation}

Comparison of the asymptotic behavior of the function $\Delta\Gamma\left(
\Lambda\right)  $ for small and large $\Lambda$, Eqs. $\left(  \ref{smlam}%
\right)  -\left(  \ref{llpower}\right)  $, indicates that this function should
have a maximum at $\Lambda=\sqrt{2}R/\delta\sim1$ except for the correlators
with small $\nu$. In experiment, the position of this maximum on the frequency
dependence of the attenuation can become a direct measurement of the
correlation radius (size) of the surface inhomogeneities $R$.

\begin{figure}[ptb]
\includegraphics{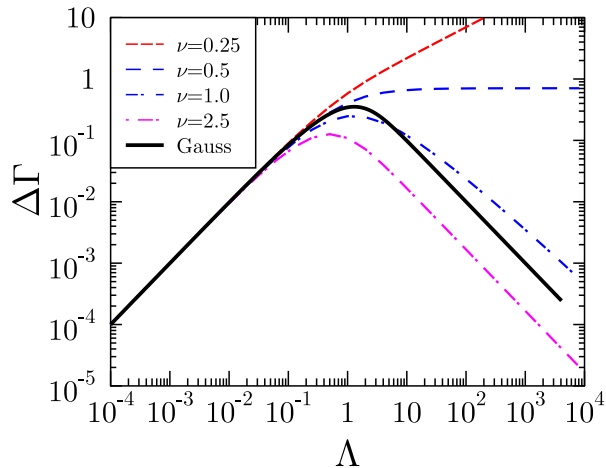}\caption{Correction to the energy dissipation rate,
$\Delta\Gamma$ as a function of the frequency parameter $\Lambda$ for Gaussian
and $\nu$-correlators in log-log scale.}%
\label{fig2}%
\end{figure}

The numerical results are summarized in Fig.~\ref{fig2} which presents the
functions $\Delta\Gamma\left(  \Lambda\right)  $ for various correlators.
Numerical values of the position of the maximum for $\Delta\Gamma\left(
\Lambda\right)  $ for various correlation functions are presented in the
Table~\ref{tab:table1}.

The last column in the Table describes the dimensionless roughness-driven
stick-slip length $\ell_{1}=\mathcal{L}_{eff}/\epsilon^{2}R$, Eq. $\left(
\ref{sl4}\right)  $, for various correlators at small $\Lambda,$
\begin{equation}
-\ell_{1}\left(  \Lambda\ll1\right)  =2\int_{0}^{\infty}\frac{dk_{x}}{\pi
}\zeta\left(  k_{x}\right)  k_{x}.\label{z3}%
\end{equation}

\section{Summary}

In summary, we calculated roughness-driven contributions to the hydrodynamic
flows, energy dissipation, and the friction force in a wide range of
parameters. We also investigated the possibility of replacing a random rough
surface by a set of effective stick-slip boundary conditions on a flat
surface. Such a replacement is highly desirable for analysis of experimental
data and/or simplification of hydrodynamic computations in microchannels. The
replacement turned out to be possible when the hydrodynamic decay length (the
viscous wave penetration depth) is larger than the size of random surface
inhomogeneities, Eq.$\left(  \ref{cond}\right)  $. The effective boundary
conditions contain two constants: the stick-slip length and the
renormalization of viscosity near the boundary. The stick-slip length and the
renormalization coefficient are expressed explicitly via the correlation
function of surface inhomogeneities. The corresponding expressions are quite
simple and can be easily used for analysis of experimental data or in
hydrodynamic computations. The effective stick-slip length is negative meaning
the effective average slow-down of the hydrodynamic flow by the rough surface
(stick rather than slip motion).

For better understanding of the results, in Appendix B below we present a
simple hydrodynamic model that illustrates our general hydrodynamic calculations.

In the process of the derivation of the effective boundary condition, we
reduced the Navier-Stokes equation with the boundary condition on a random
rough wall to the exactly equivalent closed integral equation with the
homogeneous boundary condition on the ideal flat wall. All the information on
the surface roughness is contained in the kernel of this integral equation.
The equation can be solved by standard methods.

The effective boundary parameters are analyzed numerically for three classes
of surface correlators including the Gaussian, power-law and exponentially
decaying correlators. The energy dissipation near the rough surface is
calculated as a function of frequency for these types of the correlation
functions. The position of the maximum on the frequency dependence of the
dissipation allows one to extract the correlation radius (characteristic size)
of the surface inhomogeneities directly from, for example, experiments with
torsional quartz oscillators. The stick-slip length is also evaluated
numerically for all three classes of surface correlators.

The next step should be the evaluation of the effective stick-slip length for
ultrathin flow channels of the thickness $L$ for which $L$ is expected to
gradually replace the decay length $\delta$ in the expressions for the slip
length. Another desirable development would be the incorporation into the
effective slip length of both surface and bulk scattering processes beyond the
simple Matthiessen's rule in the same spirit as recent calculations for helium
flows in microchannels \cite{mey3}.

\begin{acknowledgments}
The work is supported by NSF grant DMR-0077266.
\end{acknowledgments}

\appendix

\section{Solution of the Navier-Stokes equation for fluids restricted by
random rough walls}

First, we assume that all the variables have the harmonic time dependence,
$\exp\left(  -i\omega t\right)  $, transform the linearized Navier-Stokes
equation $\left(  \ref{e11}\right)  $ to the non-inertial coordinate frame in
which the wall is at rest,
\[
v_{x}\mathbf{\rightarrow} v_{x}\mathbf{-}\exp\left(  -i\omega t\right)
,\ x\rightarrow x-\int\exp\left(  -i\omega t\right)  dt,
\]
and introduce the stream function $\psi\left(  x,y\right)  $ as
\begin{equation}
v_{x}=\frac{\partial\psi}{\partial y},\;v_{y}=-\frac{\partial\psi}{\partial
x}.\label{eq114}%
\end{equation}
In this reference frame, the Navier-Stokes equation $\left(  \ref{e11}\right)
$ and the boundary condition $\left(  \ref{b1}\right)  $ can be rewritten as
the following equation for the stream function:
\begin{align}
-i\Lambda^{2}\nabla^{2}\psi-\nabla^{4}\psi &  =0,\label{Aeq7}\\
\frac{\partial\psi\left(  x,\epsilon\xi\left(  x\right)  \right)  }{\partial
y}  &  =1,\ \frac{\partial\psi\left(  x,\epsilon\xi\left(  x\right)  \right)
}{\partial x}=0,\label{Aeq8}\\
\psi(x,\infty)  &  =const.\label{Aeq9}%
\end{align}

The difficulty in solving Eqs. $\left(  \ref{Aeq7}\right)  -\left(
\ref{Aeq9}\right)  $ originates from the presence of a random function
$\xi\left(  x\right)  $ in the boundary condition. The next step is the
coordinate transformation
\begin{equation}
x\rightarrow x,\,y\rightarrow y-\epsilon\xi\left(  x\right)  ,\label{eq111}%
\end{equation}
that flattens the wall, making the boundary condition $\left(  \ref{eq8}%
\right)  $ simple,
\begin{equation}
\frac{\partial\psi\left(  x,0\right)  }{\partial y}=1,\ \frac{\partial
\psi\left(  x,0\right)  }{\partial x}=\epsilon\xi_{x}\left(  x\right)
,\ \psi(x,\infty)=const.\label{b3}%
\end{equation}
The change in derivatives introduces the additional term $\widehat{V}\left(
\xi,\partial_{x}\right)  \psi$ into the \emph{r.h.s} of Eq.$\left(
\ref{Aeq7}\right)  $,
\begin{equation}
-i\Lambda^{2}\nabla^{2}\psi-\nabla^{4}\psi=\widehat{V}\left(  \xi,\partial
_{x}\right)  \psi\label{eq113}%
\end{equation}
where
\begin{align}
\hat{V}  &  =\epsilon\widehat{V}_{1}+\epsilon^{2}\widehat{V}_{2}+\epsilon
^{3}\widehat{V}_{3}+\epsilon^{4}\widehat{V}_{4}\label{eq112}\\
\hat{V}_{1}\psi &  =-\left(  2\lambda^{2}\xi_{{x}}\psi_{yx}+\lambda^{2}%
\xi_{xx}\psi_{{y}}+4\xi_{xxx}\psi_{yx}+6\,\xi_{xx}\psi_{yxx}+\xi_{xxxx}%
\psi_{{y}}+2\xi_{xx}\psi_{yyy}+4\xi_{{x}}\psi_{yxxx}+4\xi_{{x}}\psi
_{yyyx}\right)  ,\nonumber\\
\hat{V}_{2}\psi &  =\left(  \lambda^{2}{\xi_{{x}}}^{2}\psi_{yy}+4\xi_{xxx}%
\xi_{{x}}\psi_{yy}+12\xi_{{x}}\xi_{xx}\psi_{yyx}+6{\xi_{{x}}}^{2}\psi
_{yyxx}+2{\xi_{{x}}}^{2}\psi_{yyyy}+3{\xi_{xx}}^{2}\psi_{yy}\right)
,\nonumber\\
\hat{V}_{3}\psi &  =-2{\xi_{{x}}}^{2}\left(  3\xi_{xx}\psi_{yyy}+2\xi_{{x}%
}\psi_{yyyx}\right)  ,\nonumber\\
\hat{V}_{4}\psi &  ={\xi_{{x}}}^{4}\psi_{yyyy},\nonumber
\end{align}
and lower indices denote the differentiation of the functions $\xi$\ and
$\psi$.

The simplicity of boundary conditions in new coordinates allows us to find the
Green's function $G(x-x^{\prime}\mathbf{,}y,y^{\prime})$ for the operator in
the \textit{l.h.s.} of Eqs. $\left(  \ref{Aeq7}\right)  $, $\left(
\ref{eq113}\right)  $. With the help of this Green's function, our initial
problem with a boundary condition on the random rough surface reduces to the
compact integral equation:%

\begin{equation}
\psi\left(  k_{x},y\right)  =\psi_{inh}\left(  k_{x},y\right)  +\int
\limits_{0}^{\infty}dy^{\prime}\,G(k_{x}\mathbf{,}y,y^{\prime})\int
\limits_{-\infty}^{\infty}\frac{dk_{x}^{\prime}}{2\pi}\widehat{V}\left(
k_{x}-k_{x}^{\prime},y^{\prime}\right)  \psi\left(  k_{x}^{\prime},y^{\prime
}\right)  ,\label{Aineq1}%
\end{equation}
where we performed the Fourier transformation in $x$-direction (in the new
coordinate frame, the geometry of the boundary is independent of $x$), and
\begin{equation}
\psi_{inh}\left(  k_{x},y\right)  =\int\limits_{-\infty}^{\infty}%
dxe^{-ik_{x}x}\psi\left(  x,y\right) \label{ap10}%
\end{equation}
is a solution of Eq. $\left(  \ref{Aeq7}\right)  $ with $\widehat{V}=0$ and
boundary conditions $\left(  \ref{Aeq8}\right)  ,\left(  \ref{Aeq9}\right)  $.
With this definition of $\psi_{inh}\left(  k_{x},y\right)  ,$ the Green's
function satisfies the homogeneous boundary conditions on the wall. Note, that
Eq. $\left(  \ref{Aineq1}\right)  $ is an \emph{exact} equivalent of our
initial problem with the random rough wall and, in principle, can be solved
for an arbitrary function $\xi\left(  x\right)  .$

The function $\psi_{inh}\left(  k_{x},y\right)  $ is determined by the
characteristic equation for the operator in \emph{l.h.s.} of Eq. $\left(
\ref{Aeq7}\right)  $
\begin{equation}
s^{4}-\left(  2k_{x}^{2}-i\Lambda^{2}\right)  s^{2}-\left(  ik_{x}^{2}%
\Lambda^{2}-k_{x}^{4}\right)  =0.\label{ap11}%
\end{equation}
This equation has four solutions:
\begin{align}
k_{y}  &  =\pm s_{1},\pm s_{2};\label{ap12}\\
s_{1}  &  =-\left|  k_{x}\right|  ,\ s_{2}=\sqrt{k_{x}^{2}-i\Lambda^{2}}%
\equiv-\alpha+i\beta,\nonumber\\
\alpha,\beta &  =\frac{1}{\sqrt{2}}\sqrt{\sqrt{k_{x}^{4}+\Lambda^{4}}\pm
k_{x}^{2}}\geq0.\nonumber
\end{align}
We are interested only in the functions $\psi_{inh}\left(  k_{x},y\right)  $
that decrease at $y\rightarrow\infty$. Therefore, the general solution of the
homogeneous Eq.$\left(  \ref{Aeq7}\right)  $ with the boundary condition
$\left(  \ref{Aeq8}\right)  $\ has the form
\begin{equation}
\psi_{inh}\left(  k_{x},y\right)  =\frac{2\pi}{i\lambda}\delta\left(
k_{x}\right)  \left[  e^{i\lambda y}-1\right]  +\frac{\epsilon\xi(k_{x}%
)}{s_{2}-s_{1}}\left[  s_{2}e^{s_{1}y}-s_{1}e^{s_{2}y}\right] \label{ap13}%
\end{equation}
and contains the contribution without $\epsilon$, $\psi_{0}\left(
k_{x},y\right)  $, and the term linear in $\epsilon$. Similar calculations
yield the Green's function:
\begin{align}
G\left(  k_{x},y,y^{\prime}\right)   &  =\frac{1}{2i\Lambda^{2}}\left[
\frac{1}{s_{2}}\left(  e^{s_{2}\left|  y-y^{\prime}\right|  }-e^{s_{2}\left(
y+y^{\prime}\right)  }\right)  -\frac{1}{s_{1}}\left(  e^{s_{1}\left|
y-y^{\prime}\right|  }-e^{s_{1}\left(  y+y^{\prime}\right)  }\right)  \right]
\label{ap14}\\
&  +\frac{1}{i\Lambda^{2}\left(  s_{2}-s_{1}\right)  }\left[  e^{s_{1}\left(
y+y^{\prime}\right)  }+e^{s_{2}\left(  y+y^{\prime}\right)  }-e^{s_{1}%
y+s_{2}y^{\prime}}-e^{s_{1}y^{\prime}+s_{2}y}\right]  .\nonumber
\end{align}
The last result can be also obtained by noticing that our Green's function is
proportional to the difference between the Green's functions for the
two-dimensional Laplace and Helmholtz equations with the same boundary
conditions:
\begin{align*}
G(\mathbf{r,r}^{\prime})  &  =\lambda^{-2}\left(  G_{L}-G_{H}\right)  ,\\
G_{L}(\mathbf{r,r}^{\prime})  &  =-\frac{1}{2\pi}\ln\left(  R_{S}%
/R_{I}\right)  ,\ G_{2}(\mathbf{r,r}^{\prime})=\frac{i}{4}\left[
H_{0}^{\left(  1\right)  }\left(  \lambda R_{S}\right)  -H_{0}^{\left(
1\right)  }\left(  \lambda R_{I}\right)  \right]  ,\\
R_{S,I}  &  =\sqrt{\left(  x-x^{\prime}\right)  ^{2}+\left(  y\mp y^{\prime
}\right)  ^{2}}.
\end{align*}

In our case of slight roughness, it is sufficient to find only the first three
terms of the expansion of the stream function $\psi$, Eq.\ $\left(
\ref{Aineq1}\right)  $, in powers of the small parameter $\epsilon$,
$\psi=\psi_{0}+\epsilon\psi_{1}+\epsilon^{2}\psi_{2}+\ldots$ Since all the
terms in the operator $\widehat{V}$ contain $\epsilon$, the only part of
$\psi$\ without $\epsilon$ is the first term in Eq. $\left(  \ref{ap13}%
\right)  $\ for $\psi_{inh}$,
\begin{equation}
\psi_{0}\left(  k_{x},y\right)  =\frac{2\pi}{i\lambda}\delta\left(
k_{x}\right)  \left[  \exp\left(  i\lambda y\right)  -1\right]  .\label{ap15}%
\end{equation}
The first order term in $\psi$ contains the remaining part of $\psi_{inh}%
$\ and the first order term in the integral $\left(  \ref{Aineq1}\right)  $
with
\[
\hat{V}_{1}\left(  k_{x},\partial_{y}\right)  \psi_{0}=-\xi\left(
k_{x}\right)  \left[  k_{x}^{2}\lambda^{2}+k_{x}^{4}\right]  e^{i\lambda y}%
\]
Integration gives
\begin{equation}
\psi_{1}\left(  k_{x,}y\right)  =\xi\left(  k_{x}\right)  \left[  e^{i\lambda
y}+\frac{i\lambda}{s_{2}-s_{1}}\left(  e^{s_{1}y}-e^{s_{2}y}\right)  \right]
.\label{ap16}%
\end{equation}

The calculation of the second order term requires straightforward integration
for much more cumbersome expressions. However, the general expression for
$\psi_{2}$ is not required for further calculations; it is sufficient to have
only the expression for $\psi_{2}$\ averaged over the surface inhomogeneities,
$\left\langle \psi_{2}\right\rangle $. The resulting expression for the stream
function contains products of the derivatives of the surface profile
$\xi^{\left(  n\right)  }\left(  x\right)  .$ These products should be
averaged over surface inhomogeneities using the definition of the correlation
function $\zeta\left(  x\right)  $, Eqs. $\left(  \ref{eq2}\right)  $
\begin{equation}
\left\langle \xi^{\left(  n\right)  }\left(  x\right)  \xi^{\left(  m\right)
}\left(  x^{\prime}\right)  \right\rangle =\left(  -1\right)  ^{m}%
\zeta^{\left(  n+m\right)  }\left(  x-x^{\prime}\right)  .\label{ap17}%
\end{equation}
In the end, after substantial cancellations that accompany the averaging,
\begin{equation}
\left\langle \psi_{2}\left(  k_{x},y\right)  \right\rangle =\delta\left(
k_{x}\right)  \int\limits_{-\infty}^{\infty}dk_{x}^{\prime}\,\zeta
(k_{x}^{\prime})\left[  \frac{s_{1}+s_{2}}{i\lambda}\left(  i\lambda
e^{i\lambda y}+s_{1}e^{s_{1}y}-s_{2}e^{s_{2}y}\right)  \right]  .\label{ap18}%
\end{equation}

Reversing the coordinate transformation of Eq. $\left(  \ref{eq111}\right)  $
and performing the related re-expansion in $\epsilon$,
\[
\left\langle v_{x}\left(  k_{x},y-\epsilon\xi\right)  \right\rangle \simeq
v_{x}^{\left(  0\right)  }\left(  y\right)  \delta\left(  k_{x}\right)
+\epsilon^{2}\left[  \left\langle v_{x}^{\left(  2\right)  }\left(
k_{x},y\right)  \right\rangle -\left\langle \xi\frac{\partial}{\partial
y}v_{x}^{\left(  1\right)  }\left(  k_{x},y\right)  \right\rangle
+\left\langle \frac{\xi^{2}}{2}\right\rangle \frac{\partial^{2}}{\partial
y^{2}}v_{x}^{\left(  0\right)  }\left(  y\right)  \delta\left(  k_{x}\right)
\right]  ,
\]
we get for the average velocity
\begin{equation}
\left\langle v_{x}\left(  x,y\right)  \right\rangle =\exp\left(  i\lambda
y\right)  \left[  1+i\lambda\varepsilon^{2}\int\limits_{0}^{\infty}%
\frac{dk_{x}}{\pi}\zeta\left(  k_{x}\right)  \left\{  s_{1}+s_{2}%
-i\lambda/2\right\}  \right]  ,\label{ap19}%
\end{equation}
where $s_{1},s_{2}$ are given by Eq. $\left(  \ref{ap12}\right)  $.

The above equations for the stream function allow one to calculate the
roughness-driven correction to the dissipation of energy and effective friction.

Time average of the bulk dissipation per unit area of the wall can be
expressed via the stream function $\psi$\ as
\begin{equation}
Q=-\frac{\eta u_{0}^{2}}{2R}\frac{1}{A}\int dV\left\langle \overline{\left(
\frac{\partial v_{i}}{\partial x_{k}}+\frac{\partial v_{k}}{\partial x_{i}%
}\right)  ^{2}}\right\rangle =-\frac{\eta u_{0}^{2}}{R}\frac{1}{A}\int
dV\left\langle 4\overline{\Psi_{xy}^{2}}+\overline{\left(  \Psi_{yy}-\Psi
_{xx}\right)  ^{2}}\right\rangle ,\label{B4a}%
\end{equation}
where $\Psi\left(  \mathbf{r,}t\right)  =\operatorname{Re}\left[  \psi\left(
\mathbf{r}\right)  e^{-i\omega t}\right]  ,$ the overline denotes the time
average over the period of oscillations, and $\left\langle ...\right\rangle $
stands for the statistical average over the random surface inhomogeneities.
The time average $\overline{\Psi_{ik}^{2}}=\frac{1}{2}\psi_{ik}\psi_{ik}%
^{\ast}$.\ 

After the coordinate transformation (\ref{eq111}), the attenuation up to the
second order term in $\epsilon$ reduces to%

\begin{align}
Q  &  =-\frac{\eta u_{0}^{2}}{2R}\int\limits_{0}^{\infty}dy\left[  Q^{\left(
0\right)  }+\epsilon^{2}Q^{\left(  2\right)  }\right]  ,\label{Qt}\\
Q^{\left(  0\right)  }  &  =\left|  \psi_{yy}^{\left(  0\right)  }\right|
^{2}=\Lambda^{2}e^{-\sqrt{2}\Lambda y},\nonumber\\
Q^{\left(  2\right)  }  &  =\left\langle \left|  \psi_{yy}^{\left(  1\right)
}+\xi_{xx}\psi_{y}^{\left(  0\right)  }-\psi_{xx}^{\left(  1\right)  }\right|
^{2}+2\left|  \xi_{x}\psi_{yy}^{\left(  0\right)  }-\psi_{xy}^{\left(
1\right)  }\right|  ^{2}+2\left|  \psi_{xy}^{\left(  1\right)  }\right|
^{2}+2\operatorname{Re}\left[  \psi_{yy}^{\left(  0\right)  \ast}\left(
\psi_{yy}^{\left(  2\right)  }+\xi_{xx}\psi_{y}^{\left(  1\right)  }\right)
\right]  \right\rangle .\nonumber
\end{align}
Finally, we get
\begin{equation}
Q=-\frac{\eta u_{0}^{2}}{2R}\frac{\Lambda}{\sqrt{2}}\left\{  1+\epsilon
^{2}\Lambda\int\frac{dk_{x}}{2\pi}\zeta\left(  k_{x}\right)  \left[
\Lambda-\sqrt{\sqrt{k_{x}^{4}+\Lambda^{4}}-k_{x}^{2}}\right]  \right\}
.\label{B5}%
\end{equation}

The friction force acting on the area unit of the surface is
\begin{align}
\mathbf{F}  &  =\frac{\eta u_{0}}{R}\mathbf{f,\ }f_{i}=-\pi_{ik}%
n_{k},\label{z1}\\
\pi_{ik}  &  =\left(  \frac{\partial v_{i}}{\partial x_{k}}+\frac{\partial
v_{k}}{\partial x_{i}}\right)  _{y=\epsilon\xi},\ \mathbf{n}=\frac{1}%
{\sqrt{1+\epsilon^{2}\xi_{x}^{2}}}\left(
\begin{array}
[c]{c}%
\epsilon\xi_{x}\\
-1
\end{array}
\right) \label{z2}%
\end{align}
Here $\mathbf{n}$ is the unit vector normal to the surface and directed out of
the liquid. The square of this force is
\begin{equation}
f^{2}=f_{x}^{2}+f_{y}^{2}=\pi_{xy}^{2}+\frac{\pi_{yy}^{2}+\epsilon^{2}%
\xi^{\prime2}\pi_{xx}^{2}}{1+\epsilon^{2}\xi^{\prime}{}^{2}},\label{B3a}%
\end{equation}
or, via the stream function,
\begin{equation}
f^{2}=\left.  \left[  \left(  \psi_{yy}-\psi_{xx}\right)  ^{2}+4\psi_{xy}%
^{2}\right]  \right|  _{y=\varepsilon\xi}.\label{B3b}%
\end{equation}
In new coordinates $\left(  \ref{eq111}\right)  $, this expression reduces to
\[
\left\langle f^{2}\right\rangle =\left\langle \left(  1+2\epsilon^{2}\xi
_{x}^{2}\right)  \psi_{yy}^{2}\left(  x,0\right)  \right\rangle .
\]
After separating the real and imaginary parts and expanding in $\epsilon$, we
finally get%

\begin{equation}
\left\langle f^{2}\right\rangle =\frac{\Lambda^{2}}{2}\left(  1+\epsilon
^{2}\int\limits_{0}^{\infty}\frac{dk_{x}}{\pi}\zeta(k_{x})\left[
\Lambda-\sqrt{\sqrt{k_{x}^{4}+\Lambda^{4}}-k_{x}^{2}}\right]  ^{2}\right)
\label{B3c}%
\end{equation}

Note, that in this problem the friction force introduced by equation $\left(
\ref{z1}\right)  $ does not determine, after integration over the surface, the
full energy dissipation. In the case of inhomogeneous rough boundaries there
is an additional dissipative contribution related to the term with pressure,
$Pn_{i}$ in the expression for the full force acting on the unit area of the
surface. If one defines the friction force not via the stress tensor, Eq.
$\left(  \ref{z1}\right)  $, but assumes the experimental definition according
to $F=-Q/u_{0}$, then the roughness-driven correction to the friction force
will be given by Eq. $\left(  \ref{B5}\right)  $ rather than by Eq. $\left(
\ref{B3c}\right)  $. Another anomaly of this problem is that one should always
take into account both, $x$ and $y$ components of the friction force.

\section{Two-layer model}

\begin{figure}[ptb]
\includegraphics{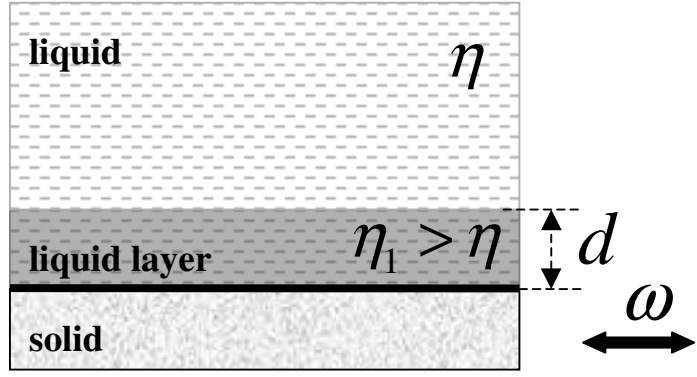}\caption{Schematic geometry of the problem.}%
\label{fig3}%
\end{figure}

The necessity of using two parameters in the effective boundary condition
instead of a single stick-slip length can be illustrated by the following
simple model. Let us consider tangential oscillations of viscous liquid which
is separated form a solid substrate by a layer of another liquid with a
slightly higher viscosity $\eta_{1}>\eta$ and the same density (see
Fig.~\ref{fig3}). In effect, we model a rough surface by a layer of viscous
liquid with somewhat different properties than in the bulk. The model has two
parameters: the thickness of the layer $d,$ and dimensionless ratio $\gamma$,
\[
\gamma=\frac{\eta\lambda}{\eta_{1}\lambda_{1}}\equiv\frac{\Lambda_{1}}%
{\Lambda}\lesssim1.
\]

Assuming that the velocity in both liquids is proportional to $\exp\left(
-i\omega t\right)  $, we get the following equations of motion:%

\begin{align*}
-i\omega v_{1}-\nu_{1}\frac{d^{2}v_{1}}{d\,y^{2}}  &  =0,\ -i\omega v-\nu
\frac{d^{2}v}{d\,y^{2}}=0,\\
v_{1}\left(  0\right)   &  =1,\ v_{1}\left(  d\right)  =v\left(  d\right)  ,\\
\eta_{1}\frac{dv_{1}\left(  d\right)  }{d\,y}  &  =\eta\frac{dv\left(
d\right)  }{d\,y}.
\end{align*}
The solution is
\begin{align*}
v_{1}\left(  y\right)   &  =Ae^{i\lambda_{1}\left(  y-d\right)  }%
+Be^{-i\lambda_{1}\left(  y-d\right)  },\\
v_{2}\left(  y\right)   &  =Ce^{i\lambda y}.
\end{align*}
with
\begin{align*}
A,B  &  =Ce^{i\lambda d}\frac{1\pm\gamma}{2},\\
C  &  =\frac{e^{-i\lambda d}}{\cos\left(  \lambda_{1}d\right)  -i\gamma
\sin\left(  \lambda_{1}d\right)  }.
\end{align*}

Time average of the rate of the energy dissipation per unit are consists of
contributions from both liquids:
\begin{align*}
Q  &  =Q_{I}+Q_{II},\\
Q_{I}  &  =-\eta_{1}\int_{0}^{d}dy\overline{\left[  \operatorname{Re}\left(
\frac{\partial v_{1}}{\partial y}e^{-i\omega t}\right)  \right]  ^{2}}\\
&  =-\left|  Ce^{i\lambda d}\right|  ^{2}\eta_{1}\Lambda_{1}\frac{1}{8\sqrt
{2}}\left[  \left(  1+\gamma\right)  ^{2}\left(  e^{\sqrt{2}\lambda_{1}%
d}-1\right)  +\left(  1-\gamma\right)  ^{2}\left(  1-e^{-\sqrt{2}\lambda_{1}%
d}\right)  +2\left(  \gamma^{2}-1\right)  \sin\left(  \sqrt{2}\lambda
_{1}d\right)  \right]  ,\\
Q_{II}  &  =-\eta\int_{d}^{\infty}dy\overline{\left[  \operatorname{Re}\left(
\frac{\partial v_{2}}{\partial y}e^{-i\omega t}\right)  \right]  ^{2}%
}=-\left|  Ce^{i\lambda d}\right|  ^{2}\frac{\eta\Lambda}{2\sqrt{2}}.
\end{align*}

If the thickness of the layer $d$ is smaller than the decay length $\delta$,
$\lambda_{1}d\ll1$,
\begin{align*}
C  &  \rightarrow1-e^{i3\pi/4}\Lambda d\left(  1-\gamma^{2}\right)  ,\\
\left|  Ce^{i\lambda d}\right|  ^{2}  &  \rightarrow1-\sqrt{2}\Lambda
d\gamma^{2}+\Lambda^{2}d^{2}\gamma^{4},\\
v\left(  y\right)   &  \approx e^{i\lambda y}\left[  1-i\lambda d\left(
1-\gamma^{2}\right)  \right] \\
-Q  &  \approx\frac{\eta\Lambda}{2\sqrt{2}}\left[  1+\Lambda^{2}d^{2}%
\gamma^{2}\left(  1-\gamma^{2}\right)  \right]  .
\end{align*}
Note, that the condition $\lambda_{1}d\ll1$ does not necessarily mean that the
layer by itself is thin.

The last two equations show that in this limit
\begin{align}
v\left(  y\geq d\right)   &  \approx\operatorname{Re}\left\{  u_{0}e^{i\left(
\lambda y-\omega t\right)  }\left[  1-e^{i3\pi/4}\Lambda d\left(  1-\gamma
^{2}\right)  \right]  \right\}  ,\label{evl1}\\
Q  &  \approx\frac{\eta u_{0}^{2}\Lambda}{2\sqrt{2}}\left[  1+\Lambda^{2}%
d^{2}\gamma^{2}\left(  1-\gamma^{2}\right)  \right]  .\label{evl2}%
\end{align}
Comparison of Eqs.$\left(  \ref{evl1}\right)  $,$\left(  \ref{evl2}\right)  $
with Eqs. $\left(  \ref{sl4}\right)  $-$\left(  \ref{sl6}\right)  $ gives the
mapping of the effective viscous layer model onto the problem with a rough
surface:
\begin{align*}
-d\left(  1-\gamma^{2}\right)   &  =\mathcal{L}_{eff}/R\equiv\epsilon^{2}%
\ell_{1},\\
d^{2}\gamma^{2}\left(  1-\gamma^{2}\right)   &  =\epsilon^{2}\ell_{2},
\end{align*}
where $\ell_{1}$, $\ell_{2}$ are given by the low-frequency equations $\left(
\ref{l1}\right)  $,$\left(  \ref{l22}\right)  $.

In this limit $d\ll\delta$, the contribution of the layer to the dissipation,
$Q_{I}$, corresponds to the $\delta-$type renormalization of the viscosity in
the effective boundary condition of Section III with renormalization parameter%
\[
\beta=\epsilon^{2}\left[  2\ell_{1}+\sqrt{2}\Lambda\ell_{2}\right]  .
\]

\bibliographystyle{apsrev}
\bibliography{slip_fv}

\end{document}